\documentclass[onecolumn,amsmath,nobibnotes,aps,prd,10pt,superscriptaddress]{revtex4-2}
\usepackage{amsmath}
\usepackage{graphicx}
\usepackage[dvipsnames]{xcolor}
\usepackage[colorlinks]{hyperref}
\hypersetup{
    linkcolor=BrickRed,
    citecolor=Green,
    urlcolor=NavyBlue
}

\usepackage{booktabs}
\usepackage{longtable}
\LTcapwidth=\textwidth

\usepackage[range-phrase=\textendash, range-units=single]{siunitx}
\let\sun\odot
\DeclareSIUnit\solarmass{\ensuremath{M_\sun}}

\begin{document}

\title{Probing neutrino emission at GeV energies from compact binary mergers with the IceCube Neutrino Observatory}
\affiliation{III. Physikalisches Institut, RWTH Aachen University, D-52056 Aachen, Germany}
\affiliation{Department of Physics, University of Adelaide, Adelaide, 5005, Australia}
\affiliation{Dept. of Physics and Astronomy, University of Alaska Anchorage, 3211 Providence Dr., Anchorage, AK 99508, USA}
\affiliation{School of Physics and Center for Relativistic Astrophysics, Georgia Institute of Technology, Atlanta, GA 30332, USA}
\affiliation{Dept. of Physics, Southern University, Baton Rouge, LA 70813, USA}
\affiliation{Dept. of Physics, University of California, Berkeley, CA 94720, USA}
\affiliation{Lawrence Berkeley National Laboratory, Berkeley, CA 94720, USA}
\affiliation{Institut f{\"u}r Physik, Humboldt-Universit{\"a}t zu Berlin, D-12489 Berlin, Germany}
\affiliation{Fakult{\"a}t f{\"u}r Physik {\&} Astronomie, Ruhr-Universit{\"a}t Bochum, D-44780 Bochum, Germany}
\affiliation{Universit{\'e} Libre de Bruxelles, Science Faculty CP230, B-1050 Brussels, Belgium}
\affiliation{Vrije Universiteit Brussel (VUB), Dienst ELEM, B-1050 Brussels, Belgium}
\affiliation{Dept. of Physics, Simon Fraser University, Burnaby, BC V5A 1S6, Canada}
\affiliation{Department of Physics and Laboratory for Particle Physics and Cosmology, Harvard University, Cambridge, MA 02138, USA}
\affiliation{Dept. of Physics, Massachusetts Institute of Technology, Cambridge, MA 02139, USA}
\affiliation{Dept. of Physics and The International Center for Hadron Astrophysics, Chiba University, Chiba 263-8522, Japan}
\affiliation{Department of Physics, Loyola University Chicago, Chicago, IL 60660, USA}
\affiliation{Dept. of Physics and Astronomy, University of Canterbury, Private Bag 4800, Christchurch, New Zealand}
\affiliation{Dept. of Physics, University of Maryland, College Park, MD 20742, USA}
\affiliation{Dept. of Astronomy, Ohio State University, Columbus, OH 43210, USA}
\affiliation{Dept. of Physics and Center for Cosmology and Astro-Particle Physics, Ohio State University, Columbus, OH 43210, USA}
\affiliation{Niels Bohr Institute, University of Copenhagen, DK-2100 Copenhagen, Denmark}
\affiliation{Dept. of Physics, TU Dortmund University, D-44221 Dortmund, Germany}
\affiliation{Dept. of Physics and Astronomy, Michigan State University, East Lansing, MI 48824, USA}
\affiliation{Dept. of Physics, University of Alberta, Edmonton, Alberta, T6G 2E1, Canada}
\affiliation{Erlangen Centre for Astroparticle Physics, Friedrich-Alexander-Universit{\"a}t Erlangen-N{\"u}rnberg, D-91058 Erlangen, Germany}
\affiliation{Physik-department, Technische Universit{\"a}t M{\"u}nchen, D-85748 Garching, Germany}
\affiliation{D{\'e}partement de physique nucl{\'e}aire et corpusculaire, Universit{\'e} de Gen{\`e}ve, CH-1211 Gen{\`e}ve, Switzerland}
\affiliation{Dept. of Physics and Astronomy, University of Gent, B-9000 Gent, Belgium}
\affiliation{Dept. of Physics and Astronomy, University of California, Irvine, CA 92697, USA}
\affiliation{Karlsruhe Institute of Technology, Institute for Astroparticle Physics, D-76021 Karlsruhe, Germany}
\affiliation{Karlsruhe Institute of Technology, Institute of Experimental Particle Physics, D-76021 Karlsruhe, Germany}
\affiliation{Dept. of Physics, Engineering Physics, and Astronomy, Queen's University, Kingston, ON K7L 3N6, Canada}
\affiliation{Department of Physics {\&} Astronomy, University of Nevada, Las Vegas, NV 89154, USA}
\affiliation{Nevada Center for Astrophysics, University of Nevada, Las Vegas, NV 89154, USA}
\affiliation{Dept. of Physics and Astronomy, University of Kansas, Lawrence, KS 66045, USA}
\affiliation{Centre for Cosmology, Particle Physics and Phenomenology - CP3, Universit{\'e} catholique de Louvain, Louvain-la-Neuve, Belgium}
\affiliation{Department of Physics, Mercer University, Macon, GA 31207-0001, USA}
\affiliation{Dept. of Astronomy, University of Wisconsin{\textemdash}Madison, Madison, WI 53706, USA}
\affiliation{Dept. of Physics and Wisconsin IceCube Particle Astrophysics Center, University of Wisconsin{\textemdash}Madison, Madison, WI 53706, USA}
\affiliation{Institute of Physics, University of Mainz, Staudinger Weg 7, D-55099 Mainz, Germany}
\affiliation{Department of Physics, Marquette University, Milwaukee, WI 53201, USA}
\affiliation{Institut f{\"u}r Kernphysik, Universit{\"a}t M{\"u}nster, D-48149 M{\"u}nster, Germany}
\affiliation{Bartol Research Institute and Dept. of Physics and Astronomy, University of Delaware, Newark, DE 19716, USA}
\affiliation{Dept. of Physics, Yale University, New Haven, CT 06520, USA}
\affiliation{Columbia Astrophysics and Nevis Laboratories, Columbia University, New York, NY 10027, USA}
\affiliation{Dept. of Physics, University of Oxford, Parks Road, Oxford OX1 3PU, United Kingdom}
\affiliation{Dipartimento di Fisica e Astronomia Galileo Galilei, Universit{\`a} Degli Studi di Padova, I-35122 Padova PD, Italy}
\affiliation{Dept. of Physics, Drexel University, 3141 Chestnut Street, Philadelphia, PA 19104, USA}
\affiliation{Physics Department, South Dakota School of Mines and Technology, Rapid City, SD 57701, USA}
\affiliation{Dept. of Physics, University of Wisconsin, River Falls, WI 54022, USA}
\affiliation{Dept. of Physics and Astronomy, University of Rochester, Rochester, NY 14627, USA}
\affiliation{Department of Physics and Astronomy, University of Utah, Salt Lake City, UT 84112, USA}
\affiliation{Dept. of Physics, Chung-Ang University, Seoul 06974, Republic of Korea}
\affiliation{Oskar Klein Centre and Dept. of Physics, Stockholm University, SE-10691 Stockholm, Sweden}
\affiliation{Dept. of Physics and Astronomy, Stony Brook University, Stony Brook, NY 11794-3800, USA}
\affiliation{Dept. of Physics, Sungkyunkwan University, Suwon 16419, Republic of Korea}
\affiliation{Institute of Physics, Academia Sinica, Taipei, 11529, Taiwan}
\affiliation{Dept. of Physics and Astronomy, University of Alabama, Tuscaloosa, AL 35487, USA}
\affiliation{Dept. of Astronomy and Astrophysics, Pennsylvania State University, University Park, PA 16802, USA}
\affiliation{Dept. of Physics, Pennsylvania State University, University Park, PA 16802, USA}
\affiliation{Dept. of Physics and Astronomy, Uppsala University, Box 516, SE-75120 Uppsala, Sweden}
\affiliation{Dept. of Physics, University of Wuppertal, D-42119 Wuppertal, Germany}
\affiliation{Deutsches Elektronen-Synchrotron DESY, Platanenallee 6, D-15738 Zeuthen, Germany}

\author{R. Abbasi}
\affiliation{Department of Physics, Loyola University Chicago, Chicago, IL 60660, USA}
\author{M. Ackermann}
\affiliation{Deutsches Elektronen-Synchrotron DESY, Platanenallee 6, D-15738 Zeuthen, Germany}
\author{J. Adams}
\affiliation{Dept. of Physics and Astronomy, University of Canterbury, Private Bag 4800, Christchurch, New Zealand}
\author{S. K. Agarwalla}
\thanks{also at Institute of Physics, Sachivalaya Marg, Sainik School Post, Bhubaneswar 751005, India}
\affiliation{Dept. of Physics and Wisconsin IceCube Particle Astrophysics Center, University of Wisconsin{\textemdash}Madison, Madison, WI 53706, USA}
\author{J. A. Aguilar}
\affiliation{Universit{\'e} Libre de Bruxelles, Science Faculty CP230, B-1050 Brussels, Belgium}
\author{M. Ahlers}
\affiliation{Niels Bohr Institute, University of Copenhagen, DK-2100 Copenhagen, Denmark}
\author{J.M. Alameddine}
\affiliation{Dept. of Physics, TU Dortmund University, D-44221 Dortmund, Germany}
\author{N. M. Amin}
\affiliation{Bartol Research Institute and Dept. of Physics and Astronomy, University of Delaware, Newark, DE 19716, USA}
\author{K. Andeen}
\affiliation{Department of Physics, Marquette University, Milwaukee, WI 53201, USA}
\author{C. Arg{\"u}elles}
\affiliation{Department of Physics and Laboratory for Particle Physics and Cosmology, Harvard University, Cambridge, MA 02138, USA}
% ...
\date{\today}

\collaboration{IceCube Collaboration}
\noaffiliation

\begin{abstract}
The advent of multi-messenger astronomy has allowed for new types of source searches by neutrino detectors. We present the results of the search for \SIrange{0.5}{100}{\GeV} astrophysical neutrinos detected with IceCube and emitted from compact binary mergers detected by the LIGO, Virgo, and KAGRA interferometers from their first run of observation (O1) to the end of the first part of the fourth (O4a). An innovative approach is used to lower the energy threshold to $\SI{0.5}{\GeV}$ and to search for an excess of GeV neutrinos in time coincidence with astrophysical transient events. Furthermore, we use a statistical combination of all observations, a binomial test, to search for a subpopulation of neutrino emitters. No significant excess was found from the studied mergers, with a best post-trial $p$-value of $40\%$, and there is currently no hint of a population of GeV neutrino emitters found in the IceCube data (post-trial $p$-value = $81\%$).
\end{abstract}

\maketitle

\section{Introduction}
\label{sec:intro}

Following the first direct detection of gravitational waves (GWs)~\cite{LIGOScientific:2016aoc}, the network of the Advanced LIGO~\cite{LIGOScientific:2014pky}, Advanced Virgo~\cite{VIRGO:2014yos}, and KAGRA~\cite{KAGRA:2020tym} detectors, operated by the LIGO, Virgo, and KAGRA collaborations (LVK), have been operating intermittently through four observing runs~\cite{LIGOScientific:2018mvr,LIGOScientific:2020ibl,LIGOScientific:2021usb,KAGRA:2021vkt}. We focus on the events observed between September 2015 (O1) and January 2024 (O4a), corresponding to about thirty months of observation time, while the remainder of the fourth observing run is not considered in this publication. These detections consist of different types of GW sources: binary black hole mergers (BBH), binary neutron star mergers (BNS), and neutron star-black hole mergers (NSBH).

The astroparticle and astrophysics communities have done follow-up observations for all reported GW candidates to identify potential counterparts in other messengers. The observation of an electromagnetic or neutrino signal would provide a multi-messenger picture of the astrophysical source, which can help constrain its physical parameters, test acceleration mechanisms, and probe the surrounding environment. The BNS merger GW170817 on August 17th, 2017~\cite{LIGOScientific:2017vwq}, was also independently observed as a Gamma-Ray Burst (GRB)~\cite{LIGOScientific:2017zic} and, through a follow-up campaign, in numerous wavelengths of the electromagnetic spectrum~\cite{LIGOScientific:2017ync}. This coincidence allowed us to understand the possible progenitors of GRB populations.

There have already been searches for neutrino counterparts, with neutrinos in the TeV--PeV range~\cite{ANTARES:2017bia, IceCube:2020xks, IceCube:2022mma, ANTARES:2018bmu}, where the atmospheric neutrino background, produced by the interaction of cosmic rays in the atmosphere, drops below the astrophysical neutrino contribution. There have also been a few searches that have reported upper limits in the single and multi-GeV range~\cite{IceCube:2023atb,KM3NeT:2023cdr,Super-Kamiokande:2021dav}.

The detection of GW170817 indicates that some of the observed merger population can be associated with electromagnetically-detected short GRBs. In such GRBs, neutrino production is possible in different energy ranges, depending on the production process. While TeV neutrinos are predicted as a consequence of internal shocks in the prompt phase of GRBs with proton-photon interactions~\cite{Halzen:2002pg}, GeV neutrinos can be produced by proton-proton and proton-neutron collisions from neutron decoupling during the acceleration phase~\cite{Murase:2013hh,Murase:2022}. The detection of GeV neutrinos from merger events or GRBs would not only offer evidence of acceleration mechanisms, but also probe the constraints of the source environment, density, and acceleration processes~\cite{Bartos:2012sg,Kimura:2018vvz,Razzaque:2003uv}. Furthermore, the ejecta of mergers can be a neutrino source by itself. The relativistic outflows can contain protons and neutrons as well, which can interact, creating GeV neutrinos \cite{Kimura:2018vvz,Murase:2013hh}. Binary black hole mergers in an active galactic nucleus accretion disk especially show promising prospects for neutrino emission~\cite{McKernan:2019hqs, Kimura:2021xxu}.

This paper reports on a search for \SIrange{0.5}{100}{\GeV} neutrinos, using the IceCube Neutrino Observatory (IceCube), from compact binary mergers detected by the LVK interferometers during the first four runs of observations. Here the mergers from the first three observation runs are taken from the Gravitational-Wave Transient Catalog (GWTC)~\cite{LIGOScientific:2018mvr,LIGOScientific:2020ibl,LIGOScientific:2021usb,KAGRA:2021vkt,ligo_scientific_collaboration_2023_10071492}, while the fourth is taken from the GCN alerts collected by GraceDB~\cite{gracedb} during the first part of the fourth Observing run (O4a) and the preceding engineering run (ER15). The energy range of neutrinos in this search is the lowest limit at which IceCube is able to detect individual neutrino events, which brings several challenges. First, at this energy range, the atmospheric neutrino background is dominant compared to the expected astrophysical source flux. Second, there is not yet any reconstruction of the incoming neutrino direction in IceCube at these energies. These challenges are mitigated by exploiting the time coincidence with the GW detection, which allows the impact of background to be reduced over short time windows. An increase in event rate in the time period of interest with respect to the expected background rate would provide evidence for neutrino production during the merger of the compact objects.

For all events, a time window for the search of $\pm \SI{500}{\second}$ centred on the merger event time is selected. This time window is motivated by considering a broad range of possible emission models of gamma-ray bursts and their observed durations~\cite{Baret:2011tk}, and is consistent with similar searches~\cite{IceCube:2020xks, IceCube:2022mma, IceCube:2023atb}. For GW events involving at least one neutron star, an additional search for a prompt signal is performed in a $[t_{\rm GW}, t_{\rm GW} + \SI{3}{\second}]$ time window. This window is motivated by the detection of a short GRB \SI{1.7}{\second} after the first detected Binary Neutron Star merger, GW170817~\cite{LIGOScientific:2017ync}, and GRB emission models~\cite{Murase:2013hh,Murase:2022,Carpio:2023wyt}.

\section{IceCube and GeV neutrino selection}
\label{sec:icecube}

The IceCube Neutrino Observatory is a cubic-kilometer neutrino detector installed in the ice at the geographic South Pole between the depths of \SI{1450}{\meter} and \SI{2450}{\meter}~\cite{IceCube:2016zyt}. It consists of 5160 digital optical modules (DOMs) distributed along 86 strings deployed within the ice. When neutrinos undergo deep inelastic scattering in the ice inside or near the IceCube detector, the subsequent muon or electromagnetic/hadronic cascades produce Cherenkov photons that can be detected by one or more DOMs. Located at the centre of the detector, IceCube DeepCore~\cite{IceCube:2011ucd} is a sub-volume component optimised to detect lower-energy events. This sub-volume is more densely populated with vertical DOM distances of \SI{7}{\meter} instead of the standard \SI{17}{\meter} on normal IceCube strings, and horizontal string distances of $\sim \SIrange{40}{70}{\meter}$ instead of the standard \SI{125}{\meter}. Furthermore, the photomultiplier tubes (PMTs) inside these DOMs have a 35\% higher quantum efficiency compared to the standard IceCube DOMs, making it easier to detect the lower intensity signals of lower energy neutrinos.
 
\begin{figure}[hbtp]
    \centering
    \includegraphics[width=0.5\linewidth]{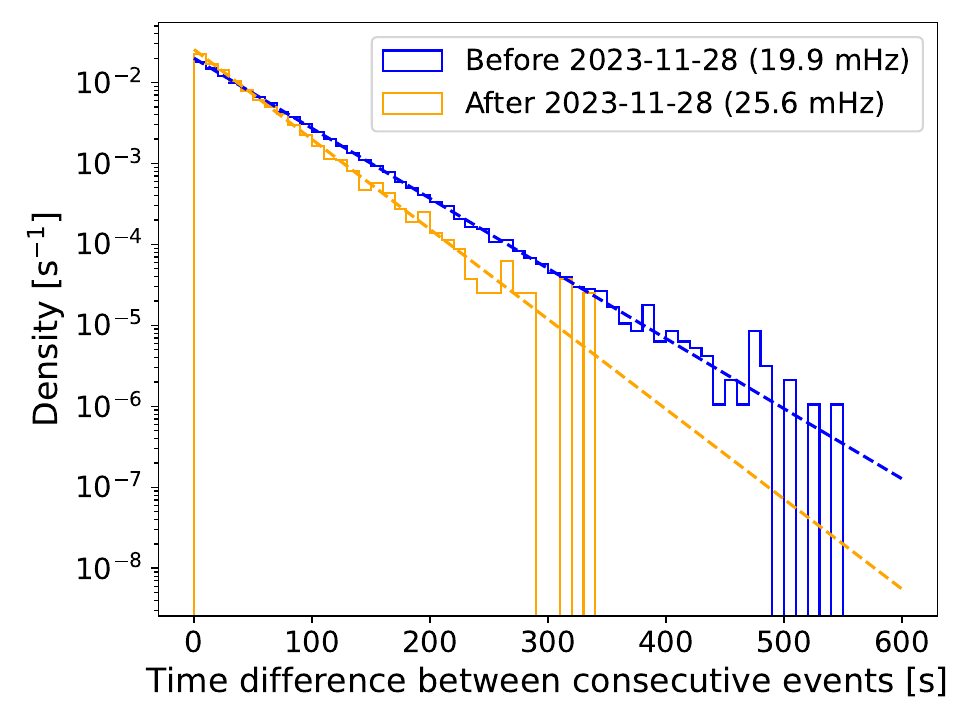}
    \caption{Histogram of the time difference between consecutive ELOWEN events, split between before and after November 28th, 2023. This date corresponds to a change in the online data processing that induced an increase in the ELOWEN rate. The y-axis shows the normalised number of occurrences per time bin and the dashed lines show the corresponding exponential fits.}
    \label{fig:timediff}
\end{figure}

\begin{figure}[hbtp]
    \centering
    \includegraphics[width=0.5\textwidth]{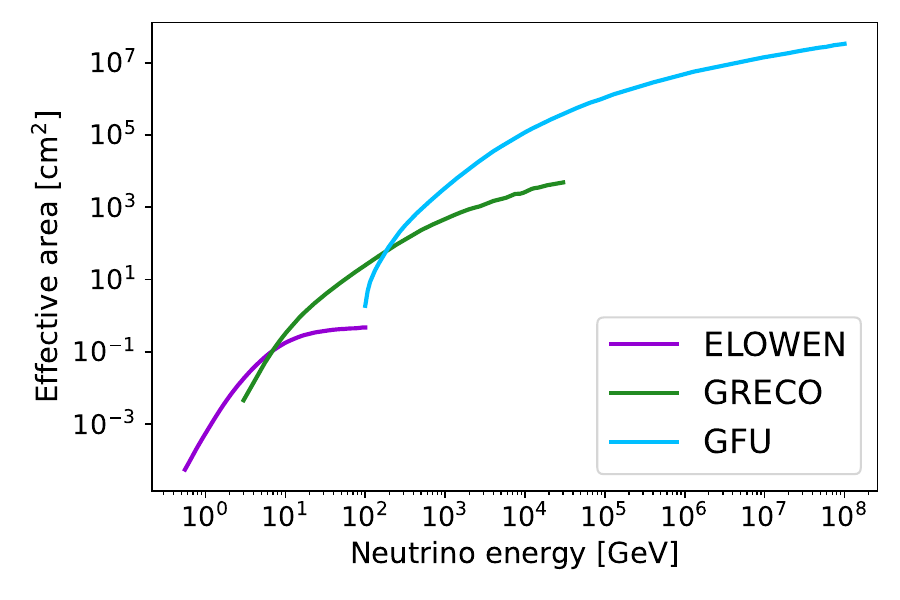}
    \caption{Effective areas averaged over the full sky as a function of neutrino energy for ELOWEN~\cite{IceCube:2021jwt}, GRECO~\cite{IceCube:2022lnv}, and GFU~\cite{IceCube:2016xci} samples, summed over neutrinos and anti-neutrinos as well as over all flavours (for GFU, only $\nu_\mu/\bar\nu_\mu$ are considered).}
    \label{fig:aeff}
\end{figure}

While IceCube was originally built to detect TeV neutrinos, with the more densely instrumented DeepCore and more sensitive trigger conditions, it is possible to go down to GeV energies~\cite{IceCube:2022lnv}, and even to detect single neutrinos with energies as low as \SI{500}{\MeV}. The ``Extremely-LOW-ENergy'' (ELOWEN) selection~\cite{IceCube:2021jwt} has been developed for such energies and is therefore employed in this article. The initial trigger rate in DeepCore~\cite{IceCube:2016zyt} is about \SI{250}{\hertz}. Successive cuts are applied to reduce the contamination from higher energy neutrinos and detector noise, as detailed in \cite{IceCube:2021jwt}. The final sample has a rate of \SI{19.9}{\milli\hertz} and retains about 40\% of the initial neutrino events simulated with energies between $0.5$ and \SI{5}{\GeV}, for all flavours, but is dominated by remaining noise. This noise consists mostly of detector noise as a mix of thermal noise and correlated noise from fluorescence or scintillation in the DOM glass \cite{Larson:2013xbf}. The contribution of atmospheric muons is minimal. Due to changes to the online data processing algorithms in the IceCube detector pertaining to its pulse extraction, the final rate after November 28th, 2023, was increased to \SI{25.6}{\milli\hertz}. The Poisson behaviour of the data and the change in rates are illustrated in \autoref{fig:timediff}, showing the time difference between consecutive events in control periods. The effective area of the ELOWEN selection can be seen in \autoref{fig:aeff}. Also shown are two higher energy event selections: the GRECO Astronomy dataset~\cite{IceCube:2022lnv}, which consists of GeV-TeV track and cascade events, and the Gamma-ray Follow-up Dataset (GFU)~\cite{IceCube:2016xci}, which consists of TeV-PeV track events. The ELOWEN selection has been used to search for neutrinos from solar flares~\cite{IceCube:2021jwt}, as well as the brightest GRB of all time, GRB221009A~\cite{IceCube:2023rhf,IceCube:2023uab}.

\section{Method}
\label{sec:method}

The ELOWEN sample is used for the search of neutrino emission in time coincidence with GW sources identified by LVK during the first four observing runs, up to the end of O4a period. Two types of analysis are performed. First, for all individual GW events, a search for neutrino counterparts is performed in a short time window around the GW detection time, the observations are converted to statistical significances and constraints on the incoming neutrino flux (\autoref{sec:method:ind}). Then, a binomial test is performed to probe the possible presence of subpopulations of sources emitting neutrinos, as described in \autoref{sec:method:pop}.

\subsection{Individual follow-ups}
\label{sec:method:ind}

Before proceeding with any further analysis, quality checks are performed to ensure the data collected by IceCube at the time of the GW event and in the preceding hours are good and usable. Notably, the detector should be operating with all the DeepCore strings, at most one or two strings are missing from the rest of the detector (to ensure proper vetoing of atmospheric muons), and the filter rates should be stable. 

Additionally, the ELOWEN rate $r_{\SI{8}{\hour}} = N_{\SI{8}{\hour}} / T_{\SI{8}{\hour}}$, where $N_{\SI{8}{\hour}}$ denotes the number of events and $T_{\SI{8}{\hour}}$ the livetime, is estimated using data from the eight hours preceding the beginning of the search time windows. This is done to account for possible small deviations due to the detector conditions at the time of the GW event, without requiring a systematic uncertainty on possible long-term evolution of the background rate. Since the ELOWEN rate is dominated by PMT noise, it is not affected by seasonal variations and $r_{\SI{8}{\hour}}$ should be consistent with the expected value, $19.9$ ($25.6$)\,\si{\milli\hertz} before (after) November 28th, 2023, within the statistical uncertainty.

If no problem has been identified, the analysis proceeds with the selection of ELOWEN events in a $\pm \SI{500}{\second}$ time window centred on the GW trigger time. The final observable, $N_{\pm \SI{500}{\second}}$, is simply the number of observed events in this window. It is compared with the expected background by computing the $p$-value as:
\begin{equation}
    p = I_{1/(1+\alpha)}(N_{\pm \SI{500}{\second}}, N_{\SI{8}{\hour}}+1),
    \label{eq:pvalue}
\end{equation}
where $I_x(a, b)$ is the regularized incomplete beta function, and $\alpha = T_{\SI{8}{\hour}} / (\SI{1000}{\second})$ is the ratio between the data livetime of the \SI{8}{\hour} time window and the duration of the search time window. As compared to a simple Poisson survival function, this estimate takes into account the statistical uncertainty on the background estimate, intrinsic to the ON/OFF approach \cite{Cousins:2007yta}. The statistical uncertainty can be estimated to be $\sim 4\%$ from approximately 600 events in the \SI{8}{\hour} time window, due the low but stable rate of ELOWEN. This approach was taken because it best suited the properties of the ELOWEN selection. A $p$-value below $0.13\%$ would correspond to a $>3\sigma$ excess with respect to the background-only hypothesis, using the one-tailed convention.

In the absence of such an excess, one may derive upper limits on the time-integrated neutrino emission from these GW sources, assuming a single power-law all-flavour $\nu+\bar\nu$ spectrum:
\begin{equation}
    F_{\rm all-flavour}^{\nu+\bar\nu}(E) = \left.\dfrac{{\rm d}N}{{\rm d}E}\right\vert_{\rm all-flavour}^{\nu+\bar\nu} = \phi \cdot \left( \dfrac{E}{\si{\giga\electronvolt}} \right)^{-\gamma}.
    \label{eq:flux}
\end{equation}
The choice of a single power law is motivated by the absence of a clear model-driven neutrino energy distribution, so that we prefer to keep the resulting constraints as model-agnostic as possible.

A Bayesian method is employed to compute the 90\% credible level (CL) upper limits on the number of signal events~\cite{Knoetig:2014dha}. The following likelihood is defined:
\begin{equation}
    \mathcal{L}(N_{\pm \SI{500}{\second}}, N_{\SI{8}{\hour}} \vert \mu_{\rm sig}, \mu_{\rm bkg}, \alpha) = \textrm{Poisson}(N_{\pm \SI{500}{\second}}; \mu_{\rm bkg} + \mu_{\rm sig}) \times \textrm{Poisson}(N_{\SI{8}{\hour}}; \alpha \mu_{\rm bkg}),
    \label{eq:lkl}
\end{equation}
where $N_{\pm \SI{500}{\second}}$ and $N_{\SI{8}{\hour}}$ are defined as before, and $\mu_{\rm sig}$ ($\mu_{\rm bkg}$) is the estimated number of signal (background) events in the search time window. A Jeffreys uninformative prior~\cite{Jeffreys:1945} is then employed for the two unknown parameters $\mu_{\rm sig}$ and $\mu_{\rm bkg}$:
\begin{equation}
    \pi(\mu_{\rm sig}, \mu_{\rm bkg}) = \left[\mu_{\rm bkg} \times (\mu_{\rm bkg} + \mu_{\rm sig})\right]^{-1/2}.
    \label{eq:prior}
\end{equation}

The posterior distribution is the product of the likelihood (\autoref{eq:lkl}) and the prior (\autoref{eq:prior}). It is then marginalised over $\mu_{\rm bkg}$ treated as a nuisance parameter, and normalized to obtain the 1D posterior $P(\mu_{\rm sig})$. The 90\% CL upper limit $\mu^{90\%}_{\rm sig}$ on the number of signal events is the unique solution of $\int_0^{\mu^{90\%}_{\rm sig}} P(\mu_{\rm sig}) {\rm d}\mu_{\rm sig} = 0.90$. This is converted to an upper limit $\phi^{90\%}$ on the flux normalisation $\phi$:
\begin{equation}
    \phi^{90\%} = \dfrac{6 \times \mu^{90\%}_{\rm sig}}{\int_{\SI{0.5}{\giga\electronvolt}}^{\SI{100}{\giga\electronvolt}} A_{\rm eff}(E) \times (E/\si{\giga\electronvolt})^{-\gamma} {\rm d}E},
\end{equation}
where $A_{\rm eff}$ is the all-flavour $\nu+\bar\nu$ ELOWEN effective area shown in \autoref{fig:aeff} and the factor $6$ accounts for the results given as the all-flavour $\nu+\bar\nu$ flux normalisation.

Finally, for GW events involving at least one neutron star, an additional search for a prompt signal was carried out in a $[t_{\rm GW}, t_{\rm GW} + \SI{3}{\second}]$ time window. For events in the GWTC catalog, the reported mass of the lighter primary source is used to determine the merger type, with events which have $\langle \mathcal{M}_2 \rangle < \SI{3}{\solarmass}$ classified by LVK as involving a neutron star. For the O4a real-time alerts, the reported probability to have a neutron star was used, with $\texttt{HasNS} > 0.5$ required. The same \SI{8}{\hour} background estimate and procedure for computing flux upper limits as for the $\pm \SI{500}{\second}$ search window are employed.

\subsection{Population study}
\label{sec:method:pop}

In addition to testing individual GW events for neutrino emission, a binomial test is performed to search for evidence of a (sub-)population of neutrino-emitting compact binary mergers. This test can uncover a set of $p$-values that are lower than expected from the background, even if none of them are individually significant. The $p$-values $p_i$ ($i=1 \dots N_\mathrm{GW}$) from the individual tests are ordered from the most to the least significant. The binomial test uses the quantity
\begin{equation}
    P(k) = \sum^N_{i=k} \frac{N!}{(N-i)!i!}p_k^i(1-p_k)^{N-1},
\end{equation}
which is the probability of having at least $k$ events with a $p$-value $p_k$ or lower, for each value of $k$. The resulting pre-trial $p$-value of the binomial test $p^\mathrm{pre}_\mathrm{binom}$ is the lowest of these $P(k)$, with a corresponding group size $k$. The post-trial $p^\mathrm{post}_\mathrm{binom}$ is then obtained as the fraction of background-only trials with that $p^\mathrm{pre}_\mathrm{binom}$ or lower, where background-only trials are generated using a Poisson distribution for each GW event and the background rate is measured from the 8-hour time window. The binomial test is performed only on the $\pm \SI{500}{\second}$ time window search.

\section{Results}
\label{sec:results}

The procedure presented in \autoref{sec:method:ind} is repeated for all GW sources reported in the GWTC (O1-O3) and published in real-time (ER15 + O4a), covering the period from September 12th, 2015, to January 16th, 2024. In the catalog, this corresponds to 93 events with high astrophysical probability ($p_{\rm astro}>50\%$). In addition, the GW200105\_162426 marginal candidate is also considered, as it is a particularly interesting neutron star -- black hole candidate merger, despite having a probability of astrophysical origin of only 36\%. For O4a, we get 85 significant real-time alerts, with a false alarm rate lower than two per year.

Out of these 179 sources, the analysis was performed for 178, excluding the S230608as alert~\cite{2023GCN.33938....1L}. The latter has not been analysed because of an IceCube charge calibration at the time of the alert. Among these, 167 sources are classified as binary black hole mergers, 9 are neutron star -- black hole merger candidates (including GW230529~\cite{LIGOScientific:2024elc}), and 2 (GW170817 and GW190425) are binary neutron star merger candidates.

\begin{figure}[hbtp]
    \centering
    \includegraphics[width=0.5\linewidth]{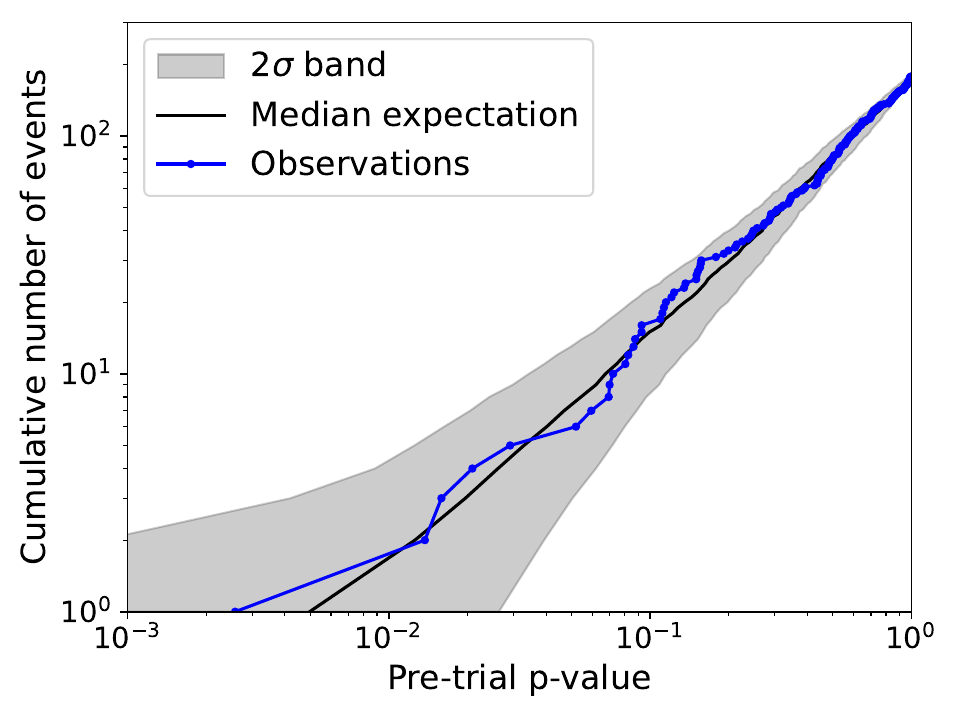}
    \caption{Cumulative distribution of the observed $p$-values for the $\pm \SI{500}{\second}$ time window search for O1-O4a GW events (blue line) compared to the background-only expectation, the black line representing the median behaviour with the corresponding $2\sigma$ band.}
    \label{fig:results}
\end{figure}

\autoref{fig:results} shows the cumulative $p$-value distribution, computed as in \autoref{eq:pvalue}, for all the events and for the $\pm \SI{500}{\second}$ time window. No significant excess with respect to the background-only hypothesis is observed. The most significant pre-trial $p$-value is $0.25\%$ for S230707ai~\cite{2023GCN.34161....1L}, corresponding to a $40\%$ post-trial $p$-value. Similarly, no event has been observed in any of the \SI{3}{\second} time windows defined for the 11 sources involving at least one neutron star. Therefore, the main outcome of the analysis is a set of upper limits on the incoming neutrino flux assuming a single-power-law flux with $\gamma = \numlist{2;2.5;3}$.

The results are reported in \autoref{tab:results1000s} for the $\pm \SI{500}{\second}$ time window and in \autoref{tab:results:3s} for the \SI{3}{\second} time window for the 11 sources involving at least one neutron star. For a $E^{-2}$ spectrum, the limits range from $E^2 F_{\rm all-flavour}^{\nu+\bar\nu}(E) =$ \SIrange{5e2}{3e3}{\giga\electronvolt\per\square\centi\meter}.

The binomial test is performed on the 178 GW events and results in a pre-trial $p$-value of $p^\mathrm{pre}_\mathrm{binom}=0.37$ for $k=30$, corresponding to a post-trial $p$-value of $p^\mathrm{post}_\mathrm{binom}=0.81$, showing no evidence of a subpopulation of GeV neutrino-emitting GW sources in the studied data set.

\section{Discussion and conclusion}

The searches carried out throughout the paper did not show any significant deviation from the background expectation, both for individual follow-ups and for the population study. The main outcomes are then upper limits on the normalisation of the incoming neutrino flux assuming a single power-law spectrum ($E^{-2}$, $E^{-2.5}$, $E^{-3}$).

\begin{figure}[hbtp]
    \centering
    \includegraphics[width=0.5\linewidth]{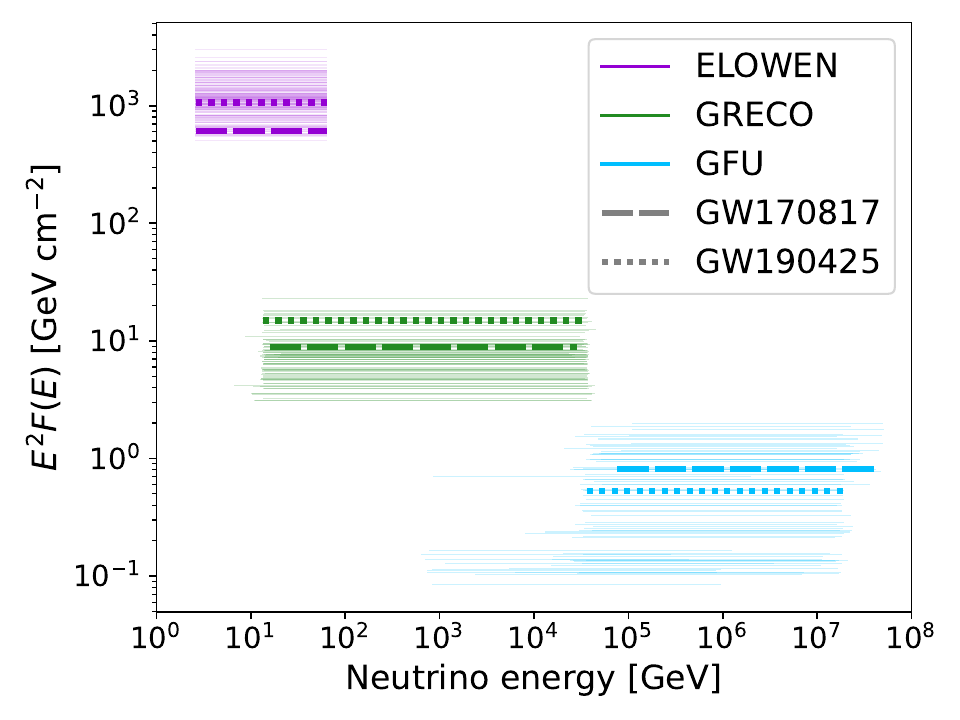}
    \caption{Summary of 90\% CL upper limits on the all-flavour time-integrated energy-squared flux $E^2 F_{\rm all-flavour}^{\nu+\bar\nu}(E)$ assuming a power-law spectrum $F_{\rm all-flavour}^{\nu+\bar\nu}(E) = \phi (E/\si{\GeV})^{-2}$. The different colours correspond to different IceCube samples: ELOWEN (this work), GRECO \cite{IceCube:2023atb}, and GFU \cite{IceCube:2022mma}. The spans in the horizontal axis correspond to the central 90\% sensitive energy range for each GW event and sample. The thicker dashed and dotted lines are the limits corresponding to the two confirmed BNS mergers, GW170817 and GW190425.}
    \label{fig:flux}
\end{figure}

These limits can be directly compared with related IceCube searches at higher energies, such as the one performed with the GRECO \cite{IceCube:2023atb} and GFU \cite{IceCube:2020xks, IceCube:2022mma} datasets. \autoref{fig:flux} shows a comparison of the limits on the flux normalisation for a $E^{-2}$ spectrum, with their corresponding central 90\% energy range. In the case of ELOWEN this corresponds to an energy range of \SIrange{2.6}{64}{\giga\electronvolt} for a $E^{-2}$ spectrum (for a $E^{-2.5}$ or $E^{-3}$ spectrum this becomes \SIrange{1.7}{43}{\giga\electronvolt} or \SIrange{1.1}{26}{\giga\electronvolt}, respectively)
 
Despite the absence of any hint of a signal, the analysis shows a promising approach to constrain the neutrino emission from GW sources (and more generally any transient population) in the GeV regime. With more GW detections being made daily, and improved sensitivity to further sources (and eventually to new categories of GW sources), population studies will be important to make the best use of all available data and possibly uncover interesting features that may otherwise remain unidentified.

The search is limited by the lack of directional reconstruction for ELOWEN events and by background from optical and electronic noise. Current work focuses on improving noise suppression and introducing a basic directional reconstruction algorithm to increase sensitivity~\cite{IceCube:2023kjn}. Another limitation comes from the minimal trigger requirement in IceCube, i.e., 3 DOMs with coincident pulses within \SI{2.5}{\micro\second}, which prevents some GeV neutrino interactions from being recorded. A separate data stream in IceCube, called HitSpooling~\cite{IceCube:2016zyt}, saves all hits occurring in the detector, independent of the implemented trigger conditions. Sub-threshold neutrino interactions, lost in regular IceCube data, can thus be saved and studied. For most of the interesting GW realtime alerts, the raw untriggered data has been saved by IceCube's HitSpool system. In the future, the plan is to exploit this data to further improve the sensitivity in the GeV regime.

The future IceCube Upgrade, consisting of seven new strings deployed inside the existing IceCube volume~\cite{Ishihara:2019aao}, will improve the efficiency and reconstruction of GeV neutrinos. The Upgrade will be equipped with multi-PMT sensors, allowing more accurate directional reconstruction and better noise rejection. In addition, new algorithms are being developed in order to exploit this more complex data structure for GeV neutrino reconstruction and selection, and therefore push further down the constraints on the neutrino flux in this energy range.

\begin{acknowledgments}
The IceCube collaboration acknowledges the significant contributions to this manuscript from Karlijn Kruiswijk, Mathieu Lamoureux, and Matthias Verreecken.
The authors gratefully acknowledge the support from the following agencies and institutions:
USA {\textendash} U.S. National Science Foundation-Office of Polar Programs,
U.S. National Science Foundation-Physics Division,
U.S. National Science Foundation-EPSCoR,
U.S. National Science Foundation-Office of Advanced Cyberinfrastructure,
Wisconsin Alumni Research Foundation,
Center for High Throughput Computing (CHTC) at the University of Wisconsin{\textendash}Madison,
Open Science Grid (OSG),
Partnership to Advance Throughput Computing (PATh),
Advanced Cyberinfrastructure Coordination Ecosystem: Services {\&} Support (ACCESS),
Frontera and Ranch computing project at the Texas Advanced Computing Center,
U.S. Department of Energy-National Energy Research Scientific Computing Center,
Particle astrophysics research computing center at the University of Maryland,
Institute for Cyber-Enabled Research at Michigan State University,
Astroparticle physics computational facility at Marquette University,
NVIDIA Corporation,
and Google Cloud Platform;
Belgium {\textendash} Funds for Scientific Research (FRS-FNRS and FWO),
FWO Odysseus and Big Science programmes,
and Belgian Federal Science Policy Office (Belspo);
Germany {\textendash} Bundesministerium f{\"u}r Bildung und Forschung (BMBF),
Deutsche Forschungsgemeinschaft (DFG),
Helmholtz Alliance for Astroparticle Physics (HAP),
Initiative and Networking Fund of the Helmholtz Association,
Deutsches Elektronen Synchrotron (DESY),
and High Performance Computing cluster of the RWTH Aachen;
Sweden {\textendash} Swedish Research Council,
Swedish Polar Research Secretariat,
Swedish National Infrastructure for Computing (SNIC),
and Knut and Alice Wallenberg Foundation;
European Union {\textendash} EGI Advanced Computing for research;
Australia {\textendash} Australian Research Council;
Canada {\textendash} Natural Sciences and Engineering Research Council of Canada,
Calcul Qu{\'e}bec, Compute Ontario, Canada Foundation for Innovation, WestGrid, and Digital Research Alliance of Canada;
Denmark {\textendash} Villum Fonden, Carlsberg Foundation, and European Commission;
New Zealand {\textendash} Marsden Fund;
Japan {\textendash} Japan Society for Promotion of Science (JSPS)
and Institute for Global Prominent Research (IGPR) of Chiba University;
Korea {\textendash} National Research Foundation of Korea (NRF),
and Chung-Ang University Research Grant;
Switzerland {\textendash} Swiss National Science Foundation (SNSF).
This research has made use of data or software obtained from the Gravitational Wave Open Science Center (gwosc.org), a service of the LIGO Scientific Collaboration, the Virgo Collaboration, and KAGRA. This material is based upon work supported by NSF's LIGO Laboratory which is a major facility fully funded by the National Science Foundation.
\end{acknowledgments}

\begin{longtable}{cc|S[table-format=2.1,table-auto-round]S[table-format=2.0,table-auto-round]S[table-format=2.1,table-auto-round]|S[table-format=2.1e1,table-auto-round]S[table-format=1.1e1,table-auto-round]S[table-format=1.1e1,table-auto-round]} 
\caption{Results of the $\pm \SI{500}{\second}$ window follow-up of GW events observed during the different observing runs: O1, O2, O3a, O3b~\cite{LIGOScientific:2018mvr,LIGOScientific:2020ibl,LIGOScientific:2021usb,KAGRA:2021vkt,ligo_scientific_collaboration_2023_10071492}, and O4a~\cite{gracedb}. The first column indicates the trigger name, and the second column is the most probable merger type. The third and fourth columns report the numbers of events in the time window expected from background $\langle \mu_{\rm bkg} \rangle = N_{\SI{8}{\hour}}/\alpha$ and observed $N_{\pm \SI{500}{\second}}$. The fifth column contains the 90\% CL upper limit on the number of signal events, and the last three columns are the corresponding 90\% CL upper limits on the all-flavour flux normalization at \SI{1}{\giga\electronvolt} $\phi$ for different spectral indices.\label{tab:results1000s}} \\
    \toprule
    {} & {} \relax & {} \relax & {} \relax & {} \relax & \multicolumn{3}{c}{Upper limits on $\phi$ [\si{\per\giga\electronvolt\per\square\centi\meter}]} \\ 
    & {} \relax & {$\langle \mu_{\rm bkg} \rangle$} \relax & {$N_{\pm \SI{500}{\second}}$} \relax & {$\mu^{90\%}_{\rm sig}$} \relax & {$\gamma = 2$} \relax & {$\gamma = 2.5$} \relax & {$\gamma = 3$} \\
    \midrule
    \endfirsthead
    \caption[]{Results \textit{(continued)}}\\
    \toprule
    \endhead
    \bottomrule
    \endfoot
    \bottomrule
    \endlastfoot
GW150914 & BBH & 21.22 \relax & 23 \relax & 10.0 \relax & 1.34e+03 \relax & 4.31e+03 \relax & 1.09e+04 \relax \\ 
GW151012 & BBH & 19.86 \relax & 22 \relax & 10.1 \relax & 1.35e+03 \relax & 4.33e+03 \relax & 1.10e+04 \relax \\ 
GW151226 & BBH & 20.63 \relax & 25 \relax & 12.2 \relax & 1.63e+03 \relax & 5.24e+03 \relax & 1.33e+04 \relax \\ 
GW170104 & BBH & 22.47 \relax & 23 \relax & 9.3 \relax & 1.24e+03 \relax & 3.99e+03 \relax & 1.01e+04 \relax \\ 
GW170608 & BBH & 18.92 \relax & 29 \relax & 17.8 \relax & 2.38e+03 \relax & 7.64e+03 \relax & 1.94e+04 \relax \\ 
GW170729 & BBH & 21.08 \relax & 19 \relax & 7.3 \relax & 9.80e+02 \relax & 3.14e+03 \relax & 7.99e+03 \relax \\ 
GW170809 & BBH & 20.63 \relax & 19 \relax & 7.5 \relax & 1.01e+03 \relax & 3.22e+03 \relax & 8.19e+03 \relax \\ 
GW170814 & BBH & 19.38 \relax & 19 \relax & 8.1 \relax & 1.08e+03 \relax & 3.47e+03 \relax & 8.83e+03 \relax \\ 
GW170817 & BNS & 19.50 \relax & 12 \relax & 4.6 \relax & 6.12e+02 \relax & 1.96e+03 \relax & 4.98e+03 \relax \\ 
GW170818 & BBH & 18.89 \relax & 24 \relax & 12.6 \relax & 1.69e+03 \relax & 5.42e+03 \relax & 1.38e+04 \relax \\ 
GW170823 & BBH & 19.76 \relax & 22 \relax & 10.2 \relax & 1.36e+03 \relax & 4.36e+03 \relax & 1.11e+04 \relax \\ 
GW190403\_051519 & BBH & 21.08 \relax & 15 \relax & 5.4 \relax & 7.17e+02 \relax & 2.30e+03 \relax & 5.84e+03 \relax \\ 
GW190408\_181802 & BBH & 19.48 \relax & 13 \relax & 4.9 \relax & 6.59e+02 \relax & 2.11e+03 \relax & 5.37e+03 \relax \\ 
GW190412 & BBH & 19.41 \relax & 20 \relax & 8.8 \relax & 1.18e+03 \relax & 3.77e+03 \relax & 9.58e+03 \relax \\ 
GW190413\_052954 & BBH & 18.61 \relax & 17 \relax & 7.1 \relax & 9.56e+02 \relax & 3.06e+03 \relax & 7.79e+03 \relax \\ 
GW190413\_134308 & BBH & 20.97 \relax & 21 \relax & 8.7 \relax & 1.16e+03 \relax & 3.72e+03 \relax & 9.45e+03 \relax \\ 
GW190421\_213856 & BBH & 21.75 \relax & 14 \relax & 4.9 \relax & 6.50e+02 \relax & 2.08e+03 \relax & 5.30e+03 \relax \\ 
GW190424\_180648 & BBH & 18.96 \relax & 20 \relax & 9.1 \relax & 1.21e+03 \relax & 3.88e+03 \relax & 9.87e+03 \relax \\ 
GW190425 & BNS & 21.08 \relax & 20 \relax & 7.9 \relax & 1.06e+03 \relax & 3.41e+03 \relax & 8.66e+03 \relax \\ 
GW190426\_152155 & NSBH & 19.62 \relax & 20 \relax & 8.7 \relax & 1.16e+03 \relax & 3.72e+03 \relax & 9.46e+03 \relax \\ 
GW190426\_190642 & BBH & 19.51 \relax & 25 \relax & 13.1 \relax & 1.75e+03 \relax & 5.62e+03 \relax & 1.43e+04 \relax \\ 
GW190503\_185404 & BBH & 18.51 \relax & 22 \relax & 11.0 \relax & 1.48e+03 \relax & 4.73e+03 \relax & 1.20e+04 \relax \\ 
GW190512\_180714 & BBH & 18.68 \relax & 19 \relax & 8.5 \relax & 1.13e+03 \relax & 3.63e+03 \relax & 9.23e+03 \relax \\ 
GW190513\_205428 & BBH & 19.35 \relax & 23 \relax & 11.3 \relax & 1.52e+03 \relax & 4.86e+03 \relax & 1.24e+04 \relax \\ 
GW190514\_065416 & BBH & 20.14 \relax & 17 \relax & 6.6 \relax & 8.77e+02 \relax & 2.81e+03 \relax & 7.15e+03 \relax \\ 
GW190517\_055101 & BBH & 19.62 \relax & 23 \relax & 11.1 \relax & 1.49e+03 \relax & 4.77e+03 \relax & 1.21e+04 \relax \\ 
GW190519\_153544 & BBH & 19.93 \relax & 12 \relax & 4.5 \relax & 6.02e+02 \relax & 1.93e+03 \relax & 4.91e+03 \relax \\ 
GW190521 & BBH & 19.69 \relax & 27 \relax & 15.0 \relax & 2.00e+03 \relax & 6.42e+03 \relax & 1.63e+04 \relax \\ 
GW190521\_074359 & BBH & 20.45 \relax & 20 \relax & 8.2 \relax & 1.10e+03 \relax & 3.53e+03 \relax & 8.98e+03 \relax \\ 
GW190527\_092055 & BBH & 19.06 \relax & 13 \relax & 5.0 \relax & 6.71e+02 \relax & 2.15e+03 \relax & 5.47e+03 \relax \\ 
GW190602\_175927 & BBH & 21.11 \relax & 15 \relax & 5.4 \relax & 7.16e+02 \relax & 2.30e+03 \relax & 5.84e+03 \relax \\ 
GW190620\_030421 & BBH & 19.06 \relax & 25 \relax & 13.5 \relax & 1.80e+03 \relax & 5.78e+03 \relax & 1.47e+04 \relax \\ 
GW190630\_185205 & BBH & 20.07 \relax & 19 \relax & 7.8 \relax & 1.04e+03 \relax & 3.33e+03 \relax & 8.46e+03 \relax \\ 
GW190701\_203306 & BBH & 20.80 \relax & 28 \relax & 15.0 \relax & 2.01e+03 \relax & 6.44e+03 \relax & 1.64e+04 \relax \\ 
GW190706\_222641 & BBH & 20.24 \relax & 21 \relax & 9.1 \relax & 1.21e+03 \relax & 3.89e+03 \relax & 9.89e+03 \relax \\ 
GW190707\_093326 & BBH & 20.97 \relax & 21 \relax & 8.7 \relax & 1.16e+03 \relax & 3.72e+03 \relax & 9.46e+03 \relax \\ 
GW190708\_232457 & BBH & 21.28 \relax & 19 \relax & 7.2 \relax & 9.68e+02 \relax & 3.10e+03 \relax & 7.89e+03 \relax \\ 
GW190719\_215514 & BBH & 19.97 \relax & 18 \relax & 7.2 \relax & 9.61e+02 \relax & 3.08e+03 \relax & 7.83e+03 \relax \\ 
GW190720\_000836 & BBH & 21.04 \relax & 17 \relax & 6.3 \relax & 8.38e+02 \relax & 2.68e+03 \relax & 6.82e+03 \relax \\ 
GW190725\_174728 & BBH & 20.00 \relax & 20 \relax & 8.5 \relax & 1.13e+03 \relax & 3.64e+03 \relax & 9.24e+03 \relax \\ 
GW190727\_060333 & BBH & 21.04 \relax & 19 \relax & 7.3 \relax & 9.82e+02 \relax & 3.15e+03 \relax & 8.00e+03 \relax \\ 
GW190728\_064510 & BBH & 17.88 \relax & 27 \relax & 16.6 \relax & 2.22e+03 \relax & 7.13e+03 \relax & 1.81e+04 \relax \\ 
GW190731\_140936 & BBH & 21.61 \relax & 10 \relax & 3.8 \relax & 5.02e+02 \relax & 1.61e+03 \relax & 4.09e+03 \relax \\ 
GW190803\_022701 & BBH & 20.92 \relax & 22 \relax & 9.5 \relax & 1.26e+03 \relax & 4.05e+03 \relax & 1.03e+04 \relax \\ 
GW190805\_211137 & BBH & 20.19 \relax & 14 \relax & 5.2 \relax & 6.90e+02 \relax & 2.21e+03 \relax & 5.62e+03 \relax \\ 
GW190814 & NSBH & 19.10 \relax & 21 \relax & 9.8 \relax & 1.31e+03 \relax & 4.19e+03 \relax & 1.06e+04 \relax \\ 
GW190828\_063405 & BBH & 19.17 \relax & 11 \relax & 4.3 \relax & 5.75e+02 \relax & 1.84e+03 \relax & 4.69e+03 \relax \\ 
GW190828\_065509 & BBH & 19.13 \relax & 14 \relax & 5.4 \relax & 7.24e+02 \relax & 2.32e+03 \relax & 5.90e+03 \relax \\ 
GW190909\_114149 & BBH & 21.32 \relax & 17 \relax & 6.2 \relax & 8.26e+02 \relax & 2.65e+03 \relax & 6.73e+03 \relax \\ 
GW190910\_112807 & BBH & 19.38 \relax & 15 \relax & 5.8 \relax & 7.75e+02 \relax & 2.49e+03 \relax & 6.32e+03 \relax \\ 
GW190915\_235702 & BBH & 19.55 \relax & 22 \relax & 10.3 \relax & 1.38e+03 \relax & 4.42e+03 \relax & 1.12e+04 \relax \\ 
GW190916\_200658 & BBH & 20.90 \relax & 21 \relax & 8.7 \relax & 1.17e+03 \relax & 3.74e+03 \relax & 9.49e+03 \relax \\ 
GW190917\_114630 & NSBH & 19.83 \relax & 16 \relax & 6.1 \relax & 8.21e+02 \relax & 2.63e+03 \relax & 6.69e+03 \relax \\ 
GW190924\_021846 & BBH & 20.80 \relax & 24 \relax & 11.2 \relax & 1.49e+03 \relax & 4.79e+03 \relax & 1.22e+04 \relax \\ 
GW190925\_232845 & BBH & 20.59 \relax & 20 \relax & 8.2 \relax & 1.09e+03 \relax & 3.51e+03 \relax & 8.91e+03 \relax \\ 
GW190926\_050336 & BBH & 20.00 \relax & 16 \relax & 6.1 \relax & 8.15e+02 \relax & 2.61e+03 \relax & 6.64e+03 \relax \\ 
GW190929\_012149 & BBH & 19.72 \relax & 16 \relax & 6.2 \relax & 8.27e+02 \relax & 2.65e+03 \relax & 6.74e+03 \relax \\ 
GW190930\_133541 & BBH & 18.41 \relax & 10 \relax & 4.1 \relax & 5.48e+02 \relax & 1.76e+03 \relax & 4.46e+03 \relax \\ 
GW191103\_012549 & BBH & 21.11 \relax & 21 \relax & 8.6 \relax & 1.15e+03 \relax & 3.69e+03 \relax & 9.38e+03 \relax \\ 
GW191105\_143521 & BBH & 20.38 \relax & 19 \relax & 7.6 \relax & 1.02e+03 \relax & 3.27e+03 \relax & 8.31e+03 \relax \\ 
GW191109\_010717 & BBH & 19.76 \relax & 11 \relax & 4.2 \relax & 5.64e+02 \relax & 1.81e+03 \relax & 4.59e+03 \relax \\ 
GW191113\_071753 & BBH & 19.72 \relax & 15 \relax & 5.7 \relax & 7.62e+02 \relax & 2.44e+03 \relax & 6.21e+03 \relax \\ 
GW191126\_115259 & BBH & 19.90 \relax & 26 \relax & 13.8 \relax & 1.84e+03 \relax & 5.90e+03 \relax & 1.50e+04 \relax \\ 
GW191127\_050227 & BBH & 20.59 \relax & 21 \relax & 8.9 \relax & 1.19e+03 \relax & 3.81e+03 \relax & 9.68e+03 \relax \\ 
GW191129\_134029 & BBH & 20.14 \relax & 18 \relax & 7.1 \relax & 9.51e+02 \relax & 3.05e+03 \relax & 7.75e+03 \relax \\ 
GW191204\_110529 & BBH & 20.39 \relax & 14 \relax & 5.1 \relax & 6.85e+02 \relax & 2.20e+03 \relax & 5.58e+03 \relax \\ 
GW191204\_171526 & BBH & 19.90 \relax & 24 \relax & 11.8 \relax & 1.58e+03 \relax & 5.07e+03 \relax & 1.29e+04 \relax \\ 
GW191215\_223052 & BBH & 19.79 \relax & 21 \relax & 9.3 \relax & 1.25e+03 \relax & 4.00e+03 \relax & 1.02e+04 \relax \\ 
GW191216\_213338 & BBH & 21.06 \relax & 12 \relax & 4.3 \relax & 5.80e+02 \relax & 1.86e+03 \relax & 4.72e+03 \relax \\ 
GW191219\_163120 & NSBH & 19.93 \relax & 15 \relax & 5.6 \relax & 7.55e+02 \relax & 2.42e+03 \relax & 6.15e+03 \relax \\ 
GW191222\_033537 & BBH & 20.35 \relax & 16 \relax & 6.0 \relax & 8.00e+02 \relax & 2.57e+03 \relax & 6.52e+03 \relax \\ 
GW191230\_180458 & BBH & 20.45 \relax & 21 \relax & 9.0 \relax & 1.20e+03 \relax & 3.84e+03 \relax & 9.76e+03 \relax \\ 
GW200105\_162426 & NSBH & 19.79 \relax & 17 \relax & 6.7 \relax & 8.93e+02 \relax & 2.86e+03 \relax & 7.28e+03 \relax \\ 
GW200112\_155838 & BBH & 20.00 \relax & 20 \relax & 8.5 \relax & 1.13e+03 \relax & 3.63e+03 \relax & 9.23e+03 \relax \\ 
GW200115\_042309 & NSBH & 20.14 \relax & 27 \relax & 14.6 \relax & 1.95e+03 \relax & 6.25e+03 \relax & 1.59e+04 \relax \\ 
GW200128\_022011 & BBH & 20.10 \relax & 24 \relax & 11.7 \relax & 1.56e+03 \relax & 5.01e+03 \relax & 1.27e+04 \relax \\ 
GW200129\_065458 & BBH & 20.97 \relax & 28 \relax & 14.9 \relax & 1.99e+03 \relax & 6.38e+03 \relax & 1.62e+04 \relax \\ 
GW200202\_154313 & BBH & 20.31 \relax & 20 \relax & 8.3 \relax & 1.11e+03 \relax & 3.57e+03 \relax & 9.07e+03 \relax \\ 
GW200208\_130117 & BBH & 19.93 \relax & 20 \relax & 8.5 \relax & 1.14e+03 \relax & 3.65e+03 \relax & 9.28e+03 \relax \\ 
GW200208\_222617 & BBH & 18.51 \relax & 19 \relax & 8.6 \relax & 1.14e+03 \relax & 3.67e+03 \relax & 9.32e+03 \relax \\ 
GW200209\_085452 & BBH & 21.34 \relax & 25 \relax & 11.7 \relax & 1.56e+03 \relax & 5.01e+03 \relax & 1.27e+04 \relax \\ 
GW200210\_092254 & NSBH & 18.37 \relax & 24 \relax & 13.1 \relax & 1.75e+03 \relax & 5.60e+03 \relax & 1.42e+04 \relax \\ 
GW200216\_220804 & BBH & 20.28 \relax & 27 \relax & 14.5 \relax & 1.93e+03 \relax & 6.20e+03 \relax & 1.58e+04 \relax \\ 
GW200219\_094415 & BBH & 19.83 \relax & 28 \relax & 15.9 \relax & 2.13e+03 \relax & 6.82e+03 \relax & 1.73e+04 \relax \\ 
GW200220\_061928 & BBH & 20.07 \relax & 26 \relax & 13.6 \relax & 1.82e+03 \relax & 5.84e+03 \relax & 1.49e+04 \relax \\ 
GW200220\_124850 & BBH & 18.72 \relax & 19 \relax & 8.4 \relax & 1.13e+03 \relax & 3.62e+03 \relax & 9.20e+03 \relax \\ 
GW200224\_222234 & BBH & 20.03 \relax & 23 \relax & 10.8 \relax & 1.45e+03 \relax & 4.64e+03 \relax & 1.18e+04 \relax \\ 
GW200225\_060421 & BBH & 20.60 \relax & 18 \relax & 6.9 \relax & 9.28e+02 \relax & 2.97e+03 \relax & 7.56e+03 \relax \\ 
GW200302\_015811 & BBH & 21.11 \relax & 19 \relax & 7.3 \relax & 9.79e+02 \relax & 3.14e+03 \relax & 7.97e+03 \relax \\ 
GW200306\_093714 & BBH & 19.72 \relax & 22 \relax & 10.2 \relax & 1.36e+03 \relax & 4.37e+03 \relax & 1.11e+04 \relax \\ 
GW200308\_173609 & BBH & 19.86 \relax & 18 \relax & 7.2 \relax & 9.67e+02 \relax & 3.10e+03 \relax & 7.88e+03 \relax \\ 
GW200311\_115853 & BBH & 19.90 \relax & 21 \relax & 9.3 \relax & 1.24e+03 \relax & 3.98e+03 \relax & 1.01e+04 \relax \\ 
GW200316\_215756 & BBH & 18.92 \relax & 20 \relax & 9.1 \relax & 1.21e+03 \relax & 3.89e+03 \relax & 9.89e+03 \relax \\ 
GW200322\_091133 & BBH & 20.31 \relax & 20 \relax & 8.3 \relax & 1.11e+03 \relax & 3.57e+03 \relax & 9.06e+03 \relax \\ 
S230518h & NSBH & 18.92 \relax & 24 \relax & 12.6 \relax & 1.68e+03 \relax & 5.40e+03 \relax & 1.37e+04 \relax \\ 
S230520ae & BBH & 19.06 \relax & 22 \relax & 10.6 \relax & 1.42e+03 \relax & 4.56e+03 \relax & 1.16e+04 \relax \\ 
S230522a & BBH & 19.27 \relax & 19 \relax & 8.2 \relax & 1.09e+03 \relax & 3.50e+03 \relax & 8.89e+03 \relax \\ 
S230522n & BBH & 19.73 \relax & 17 \relax & 6.7 \relax & 8.96e+02 \relax & 2.87e+03 \relax & 7.30e+03 \relax \\ 
GW230529 & NSBH & 20.35 \relax & 19 \relax & 7.6 \relax & 1.02e+03 \relax & 3.28e+03 \relax & 8.33e+03 \relax \\ 
S230601bf & BBH & 17.67 \relax & 28 \relax & 17.9 \relax & 2.40e+03 \relax & 7.69e+03 \relax & 1.95e+04 \relax \\ 
S230605o & BBH & 19.80 \relax & 22 \relax & 10.1 \relax & 1.36e+03 \relax & 4.35e+03 \relax & 1.11e+04 \relax \\ 
S230606d & BBH & 19.06 \relax & 17 \relax & 7.0 \relax & 9.31e+02 \relax & 2.98e+03 \relax & 7.58e+03 \relax \\ 
S230609u & BBH & 19.30 \relax & 22 \relax & 10.5 \relax & 1.41e+03 \relax & 4.52e+03 \relax & 1.15e+04 \relax \\ 
S230624av & BBH & 19.26 \relax & 18 \relax & 7.5 \relax & 1.00e+03 \relax & 3.22e+03 \relax & 8.17e+03 \relax \\ 
S230627c & NSBH & 18.82 \relax & 17 \relax & 7.1 \relax & 9.45e+02 \relax & 3.03e+03 \relax & 7.70e+03 \relax \\ 
S230628ax & BBH & 19.97 \relax & 21 \relax & 9.2 \relax & 1.24e+03 \relax & 3.96e+03 \relax & 1.01e+04 \relax \\ 
S230630am & BBH & 19.27 \relax & 20 \relax & 8.9 \relax & 1.19e+03 \relax & 3.80e+03 \relax & 9.67e+03 \relax \\ 
S230630bq & BBH & 20.56 \relax & 16 \relax & 5.9 \relax & 7.92e+02 \relax & 2.54e+03 \relax & 6.46e+03 \relax \\ 
S230702an & BBH & 20.45 \relax & 14 \relax & 5.1 \relax & 6.83e+02 \relax & 2.19e+03 \relax & 5.57e+03 \relax \\ 
S230704f & BBH & 19.41 \relax & 21 \relax & 9.6 \relax & 1.28e+03 \relax & 4.10e+03 \relax & 1.04e+04 \relax \\ 
S230706ah & BBH & 20.66 \relax & 21 \relax & 8.8 \relax & 1.18e+03 \relax & 3.79e+03 \relax & 9.63e+03 \relax \\ 
S230707ai & BBH & 19.69 \relax & 34 \relax & 22.6 \relax & 3.02e+03 \relax & 9.68e+03 \relax & 2.46e+04 \relax \\ 
S230708cf & BBH & 20.14 \relax & 19 \relax & 7.7 \relax & 1.03e+03 \relax & 3.32e+03 \relax & 8.43e+03 \relax \\ 
S230708t & BBH & 18.18 \relax & 21 \relax & 10.4 \relax & 1.39e+03 \relax & 4.45e+03 \relax & 1.13e+04 \relax \\ 
S230708z & BBH & 18.47 \relax & 14 \relax & 5.6 \relax & 7.47e+02 \relax & 2.40e+03 \relax & 6.09e+03 \relax \\ 
S230709bi & BBH & 21.32 \relax & 16 \relax & 5.7 \relax & 7.65e+02 \relax & 2.45e+03 \relax & 6.23e+03 \relax \\ 
S230723ac & BBH & 19.51 \relax & 17 \relax & 6.8 \relax & 9.07e+02 \relax & 2.91e+03 \relax & 7.39e+03 \relax \\ 
S230726a & BBH & 21.39 \relax & 14 \relax & 4.9 \relax & 6.59e+02 \relax & 2.11e+03 \relax & 5.37e+03 \relax \\ 
S230729z & BBH & 21.01 \relax & 16 \relax & 5.8 \relax & 7.76e+02 \relax & 2.49e+03 \relax & 6.32e+03 \relax \\ 
S230731an & BBH & 20.69 \relax & 27 \relax & 14.1 \relax & 1.89e+03 \relax & 6.04e+03 \relax & 1.54e+04 \relax \\ 
S230802aq & BBH & 19.86 \relax & 26 \relax & 13.8 \relax & 1.85e+03 \relax & 5.92e+03 \relax & 1.50e+04 \relax \\ 
S230805x & BBH & 18.47 \relax & 23 \relax & 12.0 \relax & 1.61e+03 \relax & 5.15e+03 \relax & 1.31e+04 \relax \\ 
S230806ak & BBH & 18.85 \relax & 18 \relax & 7.7 \relax & 1.03e+03 \relax & 3.29e+03 \relax & 8.37e+03 \relax \\ 
S230807f & BBH & 20.21 \relax & 11 \relax & 4.2 \relax & 5.56e+02 \relax & 1.78e+03 \relax & 4.53e+03 \relax \\ 
S230811n & BBH & 18.65 \relax & 18 \relax & 7.8 \relax & 1.04e+03 \relax & 3.33e+03 \relax & 8.47e+03 \relax \\ 
S230814ah & BBH & 20.00 \relax & 12 \relax & 4.5 \relax & 6.00e+02 \relax & 1.92e+03 \relax & 4.89e+03 \relax \\ 
S230814r & BBH & 20.03 \relax & 16 \relax & 6.1 \relax & 8.13e+02 \relax & 2.61e+03 \relax & 6.63e+03 \relax \\ 
S230819ax & BBH & 20.33 \relax & 22 \relax & 9.8 \relax & 1.31e+03 \relax & 4.20e+03 \relax & 1.07e+04 \relax \\ 
S230820bq & BBH & 19.51 \relax & 23 \relax & 11.2 \relax & 1.50e+03 \relax & 4.81e+03 \relax & 1.22e+04 \relax \\ 
S230822bm & BBH & 21.46 \relax & 19 \relax & 7.2 \relax & 9.59e+02 \relax & 3.07e+03 \relax & 7.81e+03 \relax \\ 
S230824r & BBH & 19.90 \relax & 19 \relax & 7.8 \relax & 1.05e+03 \relax & 3.36e+03 \relax & 8.55e+03 \relax \\ 
S230825k & BBH & 20.00 \relax & 18 \relax & 7.2 \relax & 9.59e+02 \relax & 3.07e+03 \relax & 7.81e+03 \relax \\ 
S230831e & BBH & 18.65 \relax & 16 \relax & 6.5 \relax & 8.75e+02 \relax & 2.81e+03 \relax & 7.13e+03 \relax \\ 
S230904n & BBH & 20.07 \relax & 21 \relax & 9.2 \relax & 1.23e+03 \relax & 3.93e+03 \relax & 1.00e+04 \relax \\ 
S230911ae & BBH & 19.41 \relax & 19 \relax & 8.1 \relax & 1.08e+03 \relax & 3.46e+03 \relax & 8.81e+03 \relax \\ 
S230914ak & BBH & 18.09 \relax & 19 \relax & 8.8 \relax & 1.18e+03 \relax & 3.77e+03 \relax & 9.58e+03 \relax \\ 
S230919bj & BBH & 18.85 \relax & 21 \relax & 9.9 \relax & 1.33e+03 \relax & 4.26e+03 \relax & 1.08e+04 \relax \\ 
S230920al & BBH & 18.85 \relax & 18 \relax & 7.7 \relax & 1.03e+03 \relax & 3.29e+03 \relax & 8.37e+03 \relax \\ 
S230922g & BBH & 19.13 \relax & 26 \relax & 14.4 \relax & 1.93e+03 \relax & 6.19e+03 \relax & 1.57e+04 \relax \\ 
S230922q & BBH & 18.33 \relax & 21 \relax & 10.3 \relax & 1.37e+03 \relax & 4.40e+03 \relax & 1.12e+04 \relax \\ 
S230924an & BBH & 18.75 \relax & 26 \relax & 14.8 \relax & 1.97e+03 \relax & 6.33e+03 \relax & 1.61e+04 \relax \\ 
S230927be & BBH & 19.13 \relax & 17 \relax & 6.9 \relax & 9.27e+02 \relax & 2.97e+03 \relax & 7.55e+03 \relax \\ 
S230927l & BBH & 19.82 \relax & 31 \relax & 19.1 \relax & 2.56e+03 \relax & 8.20e+03 \relax & 2.09e+04 \relax \\ 
S230928cb & BBH & 21.18 \relax & 17 \relax & 6.2 \relax & 8.32e+02 \relax & 2.67e+03 \relax & 6.78e+03 \relax \\ 
S230930al & BBH & 20.80 \relax & 26 \relax & 13.0 \relax & 1.74e+03 \relax & 5.58e+03 \relax & 1.42e+04 \relax \\ 
S231001aq & BBH & 19.41 \relax & 26 \relax & 14.2 \relax & 1.90e+03 \relax & 6.08e+03 \relax & 1.55e+04 \relax \\ 
S231005ah & BBH & 20.76 \relax & 26 \relax & 13.0 \relax & 1.74e+03 \relax & 5.59e+03 \relax & 1.42e+04 \relax \\ 
S231005j & BBH & 19.48 \relax & 25 \relax & 13.1 \relax & 1.76e+03 \relax & 5.63e+03 \relax & 1.43e+04 \relax \\ 
S231008ap & BBH & 19.31 \relax & 12 \relax & 4.6 \relax & 6.16e+02 \relax & 1.97e+03 \relax & 5.02e+03 \relax \\ 
S231014r & BBH & 18.09 \relax & 20 \relax & 9.6 \relax & 1.28e+03 \relax & 4.11e+03 \relax & 1.04e+04 \relax \\ 
S231020ba & BBH & 19.13 \relax & 20 \relax & 9.0 \relax & 1.20e+03 \relax & 3.84e+03 \relax & 9.77e+03 \relax \\ 
S231020bw & BBH & 18.99 \relax & 24 \relax & 12.5 \relax & 1.68e+03 \relax & 5.38e+03 \relax & 1.37e+04 \relax \\ 
S231028bg & BBH & 18.68 \relax & 23 \relax & 11.8 \relax & 1.58e+03 \relax & 5.08e+03 \relax & 1.29e+04 \relax \\ 
S231029y & BBH & 19.83 \relax & 19 \relax & 7.9 \relax & 1.05e+03 \relax & 3.38e+03 \relax & 8.59e+03 \relax \\ 
S231102w & BBH & 19.03 \relax & 23 \relax & 11.6 \relax & 1.55e+03 \relax & 4.96e+03 \relax & 1.26e+04 \relax \\ 
S231104ac & BBH & 19.27 \relax & 27 \relax & 15.4 \relax & 2.05e+03 \relax & 6.58e+03 \relax & 1.67e+04 \relax \\ 
S231108u & BBH & 20.42 \relax & 23 \relax & 10.6 \relax & 1.41e+03 \relax & 4.53e+03 \relax & 1.15e+04 \relax \\ 
S231110g & BBH & 20.52 \relax & 20 \relax & 8.2 \relax & 1.10e+03 \relax & 3.52e+03 \relax & 8.96e+03 \relax \\ 
S231113bb & BBH & 21.08 \relax & 20 \relax & 8.0 \relax & 1.06e+03 \relax & 3.41e+03 \relax & 8.66e+03 \relax \\ 
S231113bw & BBH & 19.58 \relax & 16 \relax & 6.2 \relax & 8.32e+02 \relax & 2.67e+03 \relax & 6.78e+03 \relax \\ 
S231114n & BBH & 20.42 \relax & 13 \relax & 4.8 \relax & 6.36e+02 \relax & 2.04e+03 \relax & 5.18e+03 \relax \\ 
S231118ab & BBH & 18.89 \relax & 12 \relax & 4.7 \relax & 6.25e+02 \relax & 2.01e+03 \relax & 5.10e+03 \relax \\ 
S231118an & BBH & 19.10 \relax & 20 \relax & 9.0 \relax & 1.20e+03 \relax & 3.85e+03 \relax & 9.79e+03 \relax \\ 
S231118d & BBH & 20.42 \relax & 16 \relax & 6.0 \relax & 7.98e+02 \relax & 2.56e+03 \relax & 6.50e+03 \relax \\ 
S231119u & BBH & 19.72 \relax & 19 \relax & 7.9 \relax & 1.06e+03 \relax & 3.40e+03 \relax & 8.64e+03 \relax \\ 
S231123cg & BBH & 19.13 \relax & 22 \relax & 10.6 \relax & 1.42e+03 \relax & 4.54e+03 \relax & 1.15e+04 \relax \\ 
S231127cg & BBH & 18.26 \relax & 17 \relax & 7.3 \relax & 9.76e+02 \relax & 3.13e+03 \relax & 7.95e+03 \relax \\ 
S231129ac & BBH & 26.08 \relax & 32 \relax & 14.5 \relax & 1.94e+03 \relax & 6.22e+03 \relax & 1.58e+04 \relax \\ 
S231206ca & BBH & 25.59 \relax & 18 \relax & 5.5 \relax & 7.38e+02 \relax & 2.37e+03 \relax & 6.01e+03 \relax \\ 
S231206cc & BBH & 25.49 \relax & 24 \relax & 8.5 \relax & 1.13e+03 \relax & 3.63e+03 \relax & 9.22e+03 \relax \\ 
S231213ap & BBH & 23.92 \relax & 32 \relax & 16.3 \relax & 2.19e+03 \relax & 7.01e+03 \relax & 1.78e+04 \relax \\ 
S231223j & BBH & 23.38 \relax & 24 \relax & 9.6 \relax & 1.28e+03 \relax & 4.10e+03 \relax & 1.04e+04 \relax \\ 
S231224e & BBH & 24.69 \relax & 22 \relax & 7.6 \relax & 1.02e+03 \relax & 3.26e+03 \relax & 8.29e+03 \relax \\ 
S231226av & BBH & 24.62 \relax & 15 \relax & 4.7 \relax & 6.30e+02 \relax & 2.02e+03 \relax & 5.13e+03 \relax \\ 
S231231ag & BBH & 26.19 \relax & 26 \relax & 9.4 \relax & 1.26e+03 \relax & 4.04e+03 \relax & 1.03e+04 \relax \\ 
S240104bl & BBH & 25.14 \relax & 21 \relax & 6.9 \relax & 9.25e+02 \relax & 2.96e+03 \relax & 7.54e+03 \relax \\ 
S240107b & BBH & 24.65 \relax & 24 \relax & 8.9 \relax & 1.18e+03 \relax & 3.80e+03 \relax & 9.65e+03 \relax \\ 
S240109a & BBH & 25.45 \relax & 29 \relax & 12.2 \relax & 1.64e+03 \relax & 5.24e+03 \relax & 1.33e+04 \relax \\ 
\end{longtable}

\begin{table}[hbtp]
    \centering
    \caption{Results of the \SI{3}{\second} window follow-up of GW events involving at least one neutron star and observed during the different observing runs: O1, O2, O3a, O3b~\cite{LIGOScientific:2018mvr,LIGOScientific:2020ibl,LIGOScientific:2021usb,KAGRA:2021vkt,ligo_scientific_collaboration_2023_10071492}, and O4a~\cite{gracedb}. The first column indicates the trigger name, and the second column is the most probable merger type. The third and fourth columns report the numbers of events in the time window expected from background $\langle \mu_{\rm bkg} \rangle = N_{\SI{8}{\hour}}/\alpha $ and observed $N_{\SI{3}{\second}}$. The fifth column contains the 90\% CL upper limit on the number of signal events, and the last three columns are the corresponding 90\% CL upper limits on the all-flavour flux normalization $\phi$ for different spectral indices.\label{tab:results:3s}}
    \begin{tabular}{cc|S[table-format=2.1]S[table-format=2.0]S[table-format=2.1]|S[table-format=2.1e1,table-auto-round]S[table-format=1.1e1,table-auto-round]S[table-format=1.1e1,table-auto-round]} 
    \toprule
    {} & {} \relax & {} \relax & {} \relax & {} \relax & \multicolumn{3}{c}{Upper limits on $\phi$ [\si{\per\giga\electronvolt\per\square\centi\meter}]} \\ 
    & {} \relax & {$\langle \mu_{\rm bkg} \rangle$} \relax & {$N_{\SI{3}{\second}}$} \relax & {$\mu^{90\%}_{\rm sig}$} \relax & {$\gamma = 2$} \relax & {$\gamma = 2.5$} \relax & {$\gamma = 3$} \\
    \midrule
GW170817 & BNS & 0.04 \relax & 0 \relax & 1.5 \relax & 2.04e+02 \relax & 6.55e+02 \relax & 1.67e+03 \relax \\ 
GW190425 & BNS & 0.04 \relax & 0 \relax & 1.5 \relax & 2.05e+02 \relax & 6.58e+02 \relax & 1.67e+03 \relax \\ 
GW190426\_152155 & NSBH & 0.04 \relax & 0 \relax & 1.5 \relax & 2.05e+02 \relax & 6.56e+02 \relax & 1.67e+03 \relax \\ 
GW190814 & NSBH & 0.04 \relax & 0 \relax & 1.5 \relax & 2.04e+02 \relax & 6.55e+02 \relax & 1.66e+03 \relax \\ 
GW190917\_114630 & NSBH & 0.04 \relax & 0 \relax & 1.5 \relax & 2.05e+02 \relax & 6.56e+02 \relax & 1.67e+03 \relax \\ 
GW191219\_163120 & NSBH & 0.04 \relax & 0 \relax & 1.5 \relax & 2.05e+02 \relax & 6.56e+02 \relax & 1.67e+03 \relax \\ 
GW200105\_162426 & NSBH & 0.04 \relax & 0 \relax & 1.5 \relax & 2.05e+02 \relax & 6.56e+02 \relax & 1.67e+03 \relax \\ 
GW200115\_042309 & NSBH & 0.04 \relax & 0 \relax & 1.5 \relax & 2.05e+02 \relax & 6.57e+02 \relax & 1.67e+03 \relax \\ 
GW200210\_092254 & NSBH & 0.04 \relax & 0 \relax & 1.5 \relax & 2.04e+02 \relax & 6.53e+02 \relax & 1.66e+03 \relax \\ 
S230518h & NSBH & 0.04 \relax & 0 \relax & 1.5 \relax & 2.04e+02 \relax & 6.54e+02 \relax & 1.66e+03 \relax \\ 
S230529ay & NSBH & 0.04 \relax & 0 \relax & 1.5 \relax & 2.05e+02 \relax & 6.57e+02 \relax & 1.67e+03 \relax \\ 
    \bottomrule
    \end{tabular}
\end{table}

\bibliography{references}

\end{document}